\documentclass[conference]{IEEEtran}
\IEEEoverridecommandlockouts
\usepackage{cite}
\usepackage{amsmath,amssymb,amsfonts}
\usepackage{algorithmic}
\usepackage{graphicx}
\usepackage{textcomp}
\usepackage{xcolor}
\usepackage{comment}
\usepackage{hyperref}
\usepackage{mdframed,lipsum}
\usepackage{comment}
\usepackage{tikz}

\newcommand{\ket}[1]{\left| #1 \right\rangle}

\renewcommand\fbox{\fcolorbox{red}{white}}
\newcommand\copyrighttext{%
  \scriptsize  M. Zheng, Y. Chen, X. Yang and A. Li, "Early Exploration of a Flexible Framework for Efficient Quantum Linear Solvers in Power Systems," 2024 IEEE Power \& Energy Society General Meeting (PESGM), Seattle, WA, USA, 2024, pp. 1-5, doi: 10.1109/PESGM51994.2024.10688916. \\ \textcopyright \the\year{} IEEE. Personal use of this material is permitted. Permission from IEEE must be obtained for all other uses, including reprinting/republishing this material for advertising or promotional purposes, collecting new collected works for resale or redistribution to servers or lists, or reuse of any copyrighted component of this work in other works.}

\newcommand\copyrightnotice{%
\begin{tikzpicture}[remember picture,overlay]
\node[anchor=south,yshift=5pt] at (current page.south) {\fbox{\parbox{\dimexpr0.9\textwidth-\fboxsep-\fboxrule\relax}{\copyrighttext}}};
\end{tikzpicture}%
}

\def\BibTeX{{\rm B\kern-.05em{\sc i\kern-.025em b}\kern-.08em
    T\kern-.1667em\lower.7ex\hbox{E}\kern-.125emX}}
\begin{document}
\title{Early Exploration of a Flexible Framework for Efficient Quantum Linear Solvers in Power Systems\\
\thanks{
The Pacific Northwest National Laboratory (PNNL) is operated by Battelle for the U.S. Department of Energy (DOE) under Contract DE-AC05-76RL01830. This work was supported by DOE Office of Electricity through its Advanced Grid Modeling (AGM) program, DOE Office of Basic Energy Science (BES), National Science Foundation CAREER, and Defense Advanced Research Projects Agency. This research used resources of the National Energy Research Scientific Computing Center (NERSC), a DOE Office of Science User Facility located at Lawrence Berkeley National Laboratory. 
Emails: {yousu.chen$|$ang.li}@pnnl.gov, and {muz219$|$xiy518}@lehigh.edu.
}
}

\author{
 \IEEEauthorblockN{Muqing Zheng\IEEEauthorrefmark{1}\IEEEauthorrefmark{2}, Yousu Chen\IEEEauthorrefmark{1}, Xiu Yang\IEEEauthorrefmark{2}, Ang Li\IEEEauthorrefmark{1}}
 \IEEEauthorblockA{\IEEEauthorrefmark{1} Pacific Northwest National Laboratory, Richland, WA, 99352, USA\\
\IEEEauthorrefmark{2}Lehigh University, Bethlehem, PA 18015, USA}
}

\maketitle

\copyrightnotice

\begin{abstract}
The rapid integration of renewable energy resources presents formidable challenges in managing power grids. While advanced computing and machine learning techniques offer some solutions for accelerating grid modeling and simulation, there remain complex problems that classical computers cannot effectively address. Quantum computing, a promising technology, has the potential to fundamentally transform how we manage power systems, especially in scenarios with a higher proportion of renewable energy sources. One critical aspect is solving large-scale linear systems of equations, crucial for power system applications like power flow analysis, for which the Harrow-Hassidim-Lloyd (HHL) algorithm is a well-known quantum solution. However, HHL quantum circuits often exhibit excessive depth, making them impractical for current Noisy-Intermediate-Scale-Quantum (NISQ) devices. In this paper, we introduce a versatile framework, powered by NWQSim, that bridges the gap between power system applications and quantum linear solvers available in Qiskit. This framework empowers researchers to efficiently explore power system applications using quantum linear solvers. Through innovative gate fusion strategies, reduced circuit depth, and GPU acceleration, our simulator significantly enhances resource efficiency. Power flow case studies have demonstrated up to a eight-fold speedup compared to Qiskit Aer, all while maintaining comparable levels of accuracy.
\end{abstract}


\section{Introduction}
As nations around the world unite in the battle against climate change, clean energy and decarbonization have taken center stage. For perspective, in 2021, one-fifth of the electricity in the U.S. came from renewable sources. With eyes set on a greener future, America is aiming for $100\%$ clean electricity by 2035 and zero net emissions by 2050. California has a pathway to achieve carbon neutral by 2045\cite{b0}. Europe is on a similar trajectory, with numerous countries launching strategies to cut their greenhouse gas emissions significantly.

The energy landscape is indeed transforming. The shift is evident with the growing reliance on renewable sources like wind and solar, advanced energy-consuming devices, and other cutting-edge technologies. Taking California as an example: there's been a rapid increase in the use of wind and solar energy. By 2021, $90\%$ of its power came from these sources. And in a milestone moment on May 8, 2022, this number surged past $100\%$. Such changes has brought significant challenges to grid management as it demands a much faster, quicker responses to a much complex, dynamic, and with uncertainty not been seen before. 

High-performance-computing, as an emerging technique, can help power system engineers to gain faster computational speeds. However, there are still problems in power systems, particularly those of large scale and high granularity, are intractable to classical cluster machines due to their computational complexity, data communication overhead, or both. Some examples are listed below: 

\begin{itemize}
\item {Stochastic Optimization for Operational Planning with Renewables}: Given the variability and uncertainty of renewable energy sources like wind and solar, planning for their optimal integration and dispatch involves solving high-dimensional stochastic optimization problems.

\item {Very Large-Scale Contingency Analysis}: Assessing the grid's resilience to numerous possible contingencies (like multiple line outages) requires analyzing a combinatorial number of scenarios. It's a problem that can quickly grow beyond the capabilities of classical computing clusters.

\item {Security Constrained Unit Commitment}: Determining which power plants should be turned on or off to meet the anticipated electricity demand, while also ensuring the grid remains stable and secure against potential disturbances.
\end{itemize}
Classical cluster machines, while powerful, have limitations in terms of processing capability and data communication bandwidth. Emerging computing paradigms, like quantum computing, might offer pathways to tackle some of these intractable problems in the future.

The basic information unit in quantum computing is called qubits. An $n$-qubit quantum state can be described as a $2^n$-size complex vector in Hilbert space. Quantum state has a unique feature: at most $2^n$ number of distinguishable, i.e., orthogonal, quantum states can form a valid $n$-qubit quantum state. This property is called superposition. Another property of quantum states is entanglement. It is when several quantum states are intricately intertwined  so that any transformation on one state will affect all the other states. Both are important sources of 

\newpage
\noindent quantum advantage.
\copyrightnotice

Quantum computing holds great promise for power systems, even though some details remain unclear. Many experts are eager to start exploring its application before it is perfected. 
The areas that they are working on include AC/DC power flow, contingency analysis, state estimation, transient stability assessment, fault diagnosis, unit commitment, and facility location–allocation problem, as shown in \cite{b1} \cite{b2}, \cite{b3}.
However, there are still gaps that need to be bridged to enable power system engineers, who typically do not have a quantum computing background, to evaluate how to use quantum algorithms to accelerate their applications. To address this, we introduce a flexible framework, empowered by Northwest Quantum Simulator (NWQSim)\cite{li2020},\cite{li2021}, to bridge the gaps between power system engineers and quantum techniques.

Building upon this potential of quantum computing for power systems, it is crucial to understand the tools that can help harness this technology effectively. Quantum solvers, for instance, leverage the principles of quantum mechanics to address specific types of problems more efficiently than classical computers. Some of the main quantum solvers include:

\begin{itemize}
    \item Quantum Linear System Algorithms (QLSAs): QLSAs are used for solving linear systems of equations efficiently on a quantum computer, one of the representative algorithms is Harrow-Hassidim-Lloyd (HHL) ~\cite{hhl},
    \item Variational Quantum Eigensolver (VQE): VQE is a quantum solver used for finding the ground state energy (the smallest eigenvalue) of quantum systems,
    \item Quantum Approximate Optimization Algorithm (QAOA): QAOA is a quantum solver designed for solving  combinatorial optimization problems, such as MaxCut problems,
    \item Quantum Annealing (QA): QA currently solves the problems that can be reformulated to Ising models and quadratic unconstrained binary optimization problems.
\end{itemize}

This paper start with integrating the HHL solver into the framework due to its ability to handle  linear system and its connection to power systems. HHL has the potential to solve complex linear equations with exponential speedup compared to classical solvers, which could enable a significant computational improvement for many power system applications,such as power flow, short-circuit analysis, contingency analysis, state estimation, and voltage stability analysis~\cite{b1,b2}.

\begin{figure}[t]
    \centering
    \includegraphics[width=0.99\linewidth]{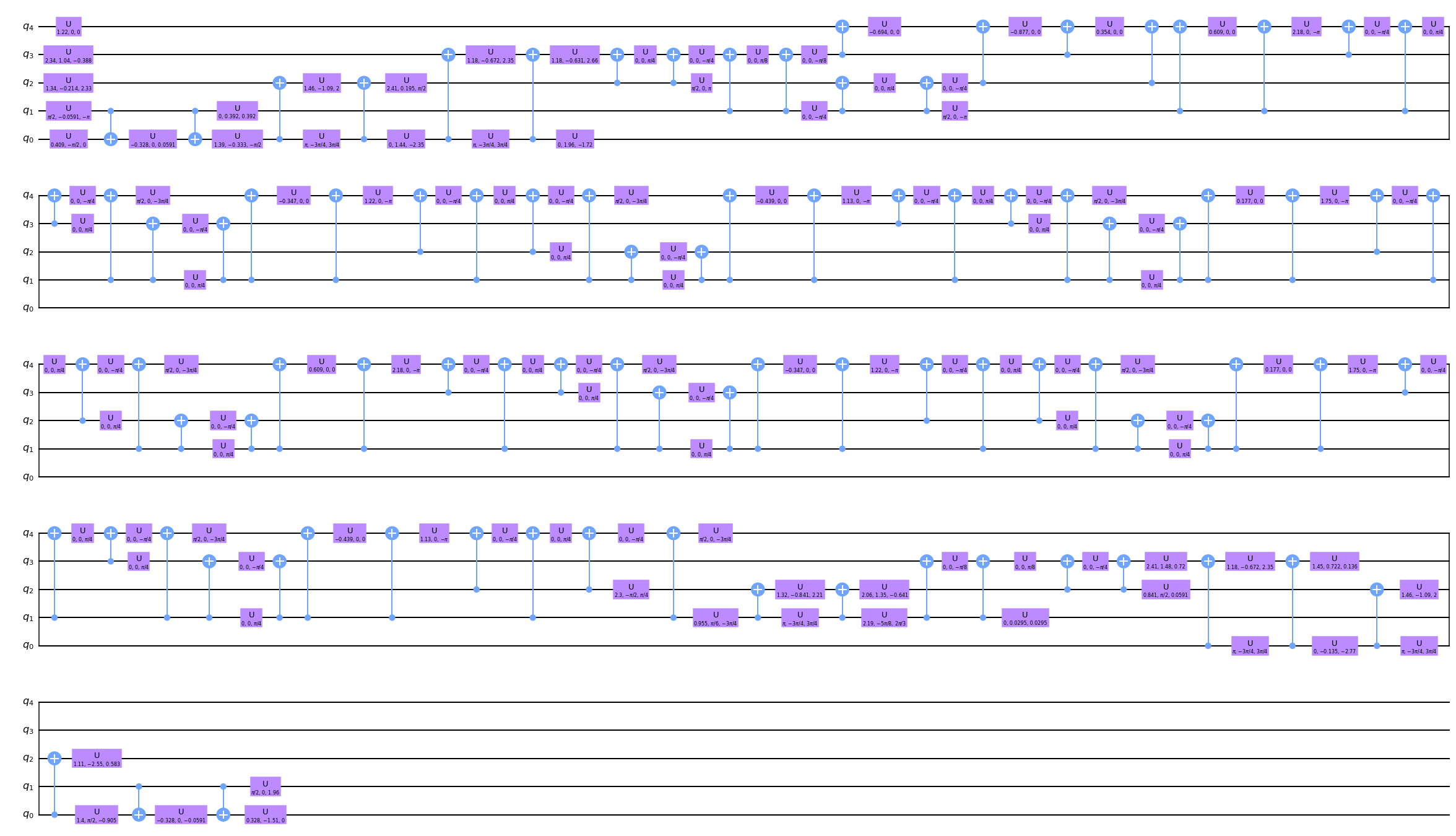}
    \caption{An example of the \textit{transpiled} HHL circuit that solves a random $2$-by-$2$ linear system. It is generated by the package in~\cite{qlsarepo} and transpiled into one- and two-qubit basis gates.}
    \label{fig:2-by-2circuit}
\end{figure}

\begin{figure}[b]
    \centering
    \includegraphics[width=0.77\linewidth]{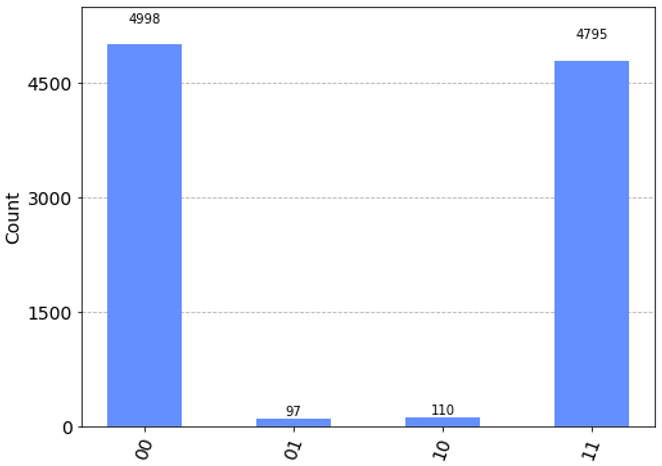}
    \caption{10,000 measurements of bell state $\frac{1}{\sqrt{2}}\left(\ket{00}+\ket{11}\right)$ prepared from initial state $\ket{00}$ using a Hadamard gate and a Controlled-NOT gate on IBM superconducting quantum computer ibmq\_kolkata. The noise in the quantum computer causes the actual results deviate from the ideal case where $01$ and $10$ should not appear and the counts for $00$ and $11$ should be almost equal in the measurements.}
    \label{fig:computer_noise}
\end{figure}

Nevertheless, applying the HHL algorithm directly to power system problems presents several challenges. First and foremost, the quantum circuits for the HHL algorithm tend to be very deep, even for small-scale problems. To illustrate, consider the HHL circuit that only solves a random $2$-by-$2$ linear system, depicted in Figure~\ref{fig:2-by-2circuit}. It already has 120 one-qubit gates and 90 two-qubit gates~\cite{qlsarepo}. Furthermore, the small experiment illustrated in Figure~\ref{fig:computer_noise} shows that significant errors persist even in a circuit with one-qubit gate and a two-qubit gate. In essence, given the current noise level in Noisy Intermediate-Scale Quantum (NISQ) devices, obtaining meaningful results for such circuits is difficult. To the best of our knowledge, a $4$-by-$4$ system is the largest one that can be accurately solved on an actual quantum computer in existing literature~\cite{yalovetzky2021nisq}, which falls short of the requirements for practical power system applications.

Additionally, HHL circuits have certain requirements on qubit connectivity. This is not an issue for trapped-ion quantum computers as discussed in~\cite{yalovetzky2021nisq}, but it is challenging for superconducting quantum computers offered by companies like IBM and Google. Lastly, there is a lack of software tools for developing and deploying quantum algorithms for power system applications, especially for non-quantum-background researchers and engineers.

Given all these challenges, the authors propose a flexible simulator-based framework designed for tackling power system problems. This framework leverages quantum simulators, which are dedicated to the evaluation of quantum circuits on classical quantum computers. The use of simulators can facilitate feasibility studies in this area, targeting more realistic problem settings and making it more accessible to a broader range of researchers. Once fault-tolerant quantum computers are available, researchers can then migrate their existing work to the real hardware, further advancing the field.

\newpage
\copyrightnotice

The remainder of this paper is organized as follows: Section~\ref{sec:into-framework} introduces the flexible framework along with its key components and setup. Section~\ref{sec:details} explains the details of applying the HHL algorithm to linear systems in power flow problems. Section~\ref{sec:num-exp} illustrates the use of our framework on three power flow problems. Finally, Section~\ref{sec:conclusion} concludes the paper with a discussion of future work.

\section{Introduction to the Framework\label{sec:into-framework}}

The motivation is to provide power system engineers with a versatile platform for testing power system applications by seamlessly combining classical and quantum algorithms within a hybrid simulation framework. This hybrid simulation framework allows users to connect with various existing power system applications with different quantum simulators via the interfaces that facilitate the smooth exchange of data necessary for the successful execution of hybrid simulations. The ultimate goal is to equip power system professionals with the tools and environment they need to explore and advance quantum computing in power system applications.

\subsection{The proposed hybrid framework}

\begin{figure}[b]
\centerline{\includegraphics[width=0.75\linewidth]{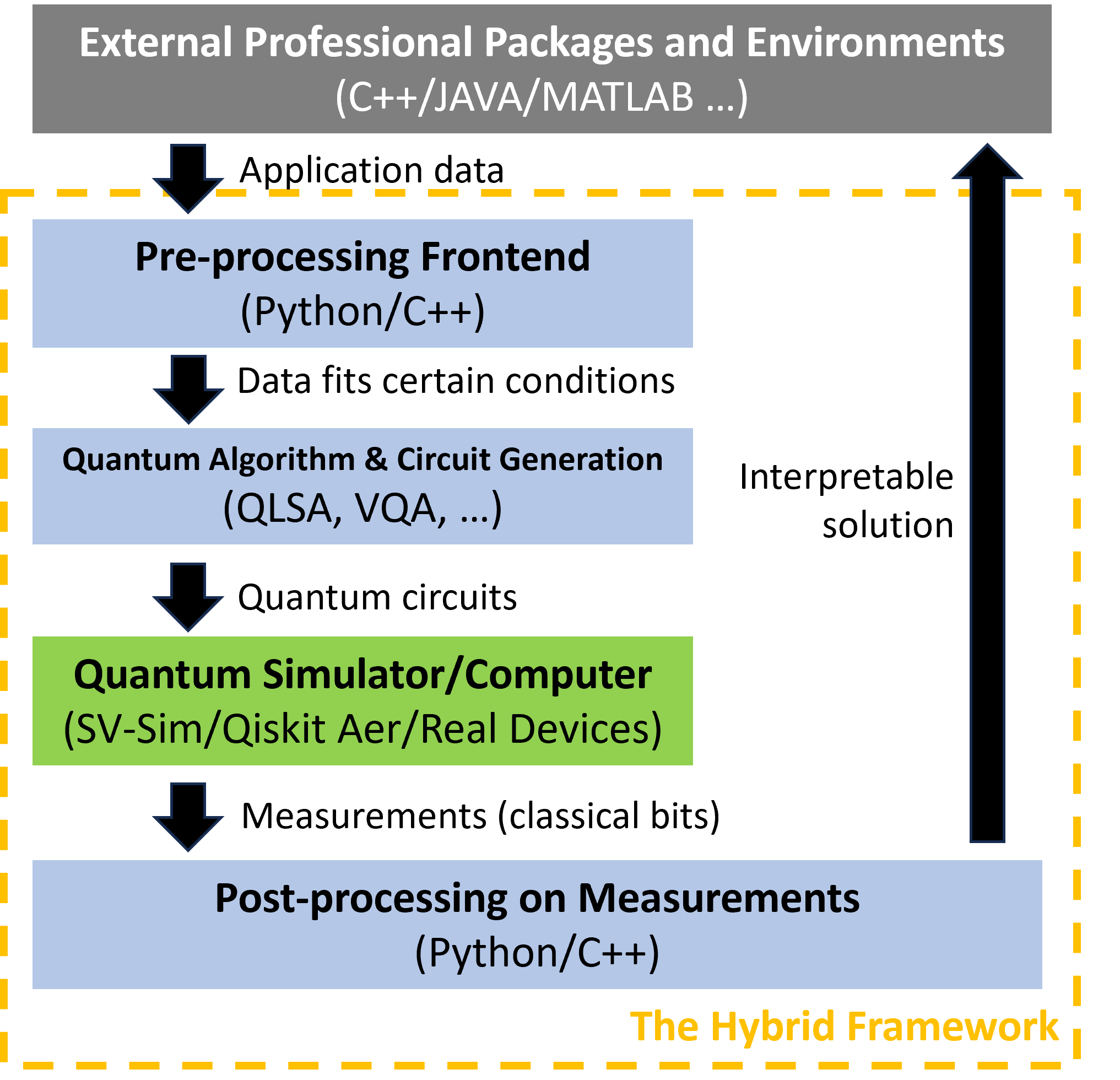}}
\caption{The conceptual view of the proposed framework.}
\label{fig:framework}
\end{figure}

As illustrated in Figure \ref{fig:framework}, our conceptual framework serves as a bridge, enhancing the communication between various quantum algorithms, such as HHL and variational quantum algorithm (VQA), and external simulation tools compatible with diverse languages and platforms. Our current focus lies in the integration of the HHL algorithm with the power flow functions in MatPower \cite{matpower}. This integration necessitates some data preprocessing to transform the data into a format suitable for HHL. Subsequently, we are poised to extend our framework by integrating other quantum algorithms, such as VQA, to further enrich our hybrid simulation capabilities.

\subsection{Existing software stack choices}

On the classical side of the framework, as a starting point, we have integrated MatPower due to its widespread use in power system problems in academia. For the quantum computing component of our framework, we have selected Qiskit\cite{qiskit} developed by IBM, from among other tools such as Cirq from Google, and Q\# from Microsoft. Qiskit offers several advantages that make it a preferred choice. Foremost is the large and active community supporting Qiskit as an open-source python library. This helps lower the potential difficulties on the development of the framework. Additionally, Qiskit provides a wide range of implementations for popular quantum algorithms, which will assist the future integration of additional quantum algorithms into the framework. Notice that Qiskit has provided a complete VQA routine, but its recent update removed its HHL module. To address this, we have adapted a Qiskit-based HHL package in~\cite{qlsarepo} and incorporated it into the framework.

Although Qiskit has provided quantum simulators in its Aer submodule, their efficiency falls short when dealing with complex circuits like those in the HHL algorithm. To overcome this limitation, our preferred simulator is the SV-Sim simulator in NWQSim. It is a C++ CPU/GPU statevector simulator includes various front-end supports such as C++, Qiskit, native Python, and Q\#. When compared to the statevector simulators in Qiskit, Cirq, and Q\#, SV-Sim exhibits exceptional performance in the benchmarking tests in~\cite{li2021}. The speed comparisons between simulators in Qiskit Aer and NWQSim on power flow use cases are also shown in Section~\ref{sec:num-exp}.

\subsection{Embedded techniques in SV-Sim simulator}

\begin{figure}[t]
\centering
\includegraphics[width=0.65\linewidth]{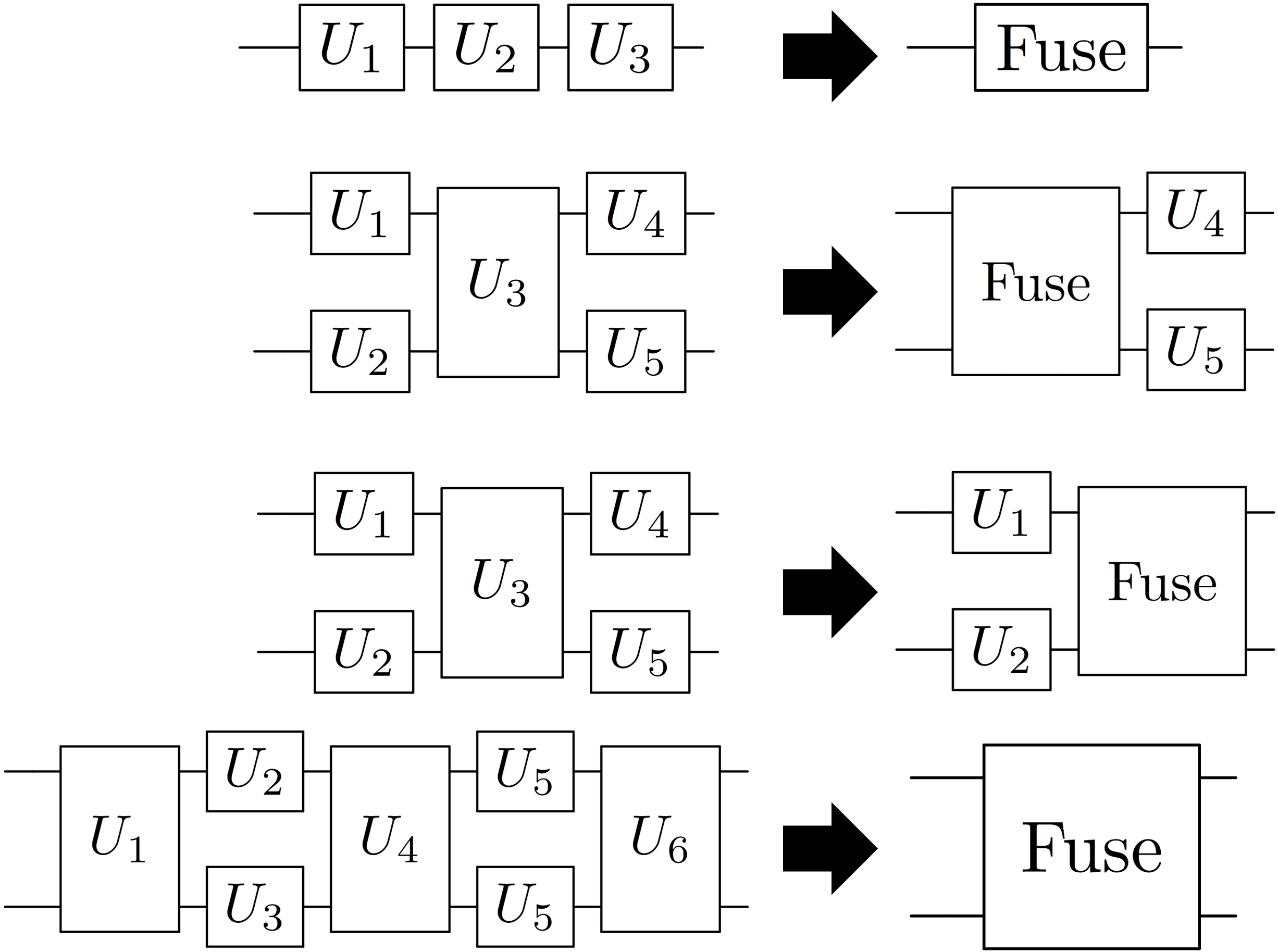}
\caption{Four possible gate fusion strategies for \textit{applicable gates}. The individual gates on the left-hand side are combined into one fused gate on the right-hand side. The priorities among four strategies is from top to the bottom: first, one-qubit gates are fused; then two different fusion between one-qubit and two-qubit gates happen; and finally, two-qubit gates are fused together. }
\label{fig:fusion}
\end{figure}

The majority of time consumption during HHL circuit simulation is attributed to the process of simulation and measurement. Therefore, it's crucial to highlight several techniques employed by SV-Sim (NWQSim) to optimize computation time for deep circuits. The most significant time sink is the unitary evolution of the state. For SV-Sim, a noticeable improvement over the simulators in Aer (from Qiskit) and qsim (from Cirq) is the specialized gate implementation.  While other simulators use general unitary gates for all one-qubit and two-qubit gates, SV-Sim offers specialized computation for dozens of supported basis gates and specific architectures of CPUs and GPUs. One instance of such specialized computation is gate fusion. As illustrated in Figure~\ref{fig:fusion}, gate fusion merges several applicable gates into a single gate based on four distinct strategies. This substantially reduces both the gate counts and the depth of the circuits. Taking the 210-gate circuit shown in Figure~\ref{fig:2-by-2circuit} as an example, the actual circuit executed in SV-Sim only has 67 gates. While gate fusion is limited to certain basis gates, it does not necessitate the matrix multiplications. Thus, the overall runtime can be lowered significantly~\cite{li2021}.


Another source of time inefficiency stems from latency arising from communication between multiple computational units during parallel computing. On one hand, SV-Sim employs the so-called ``PGAS-based SHMEM'' communication model, which provides enhanced support for software governing direct links between GPUs compared to the traditional MPI routine.

\newpage
\noindent On the other hand, researchers have also developed a novel method to minimize unnecessary data migration~\cite{li2021}.
\copyrightnotice

\section{Power Flow Use Case\label{sec:details}}

Power flow is a fundamental power system application. In this paper, our primary focus centers on the Newton-Raphson based AC power flow, a widely used technique for power system analysis. 

\subsection{Procedures for applying HHL on practical problems\label{sec:hhl-procedures}}

Let us consider the linear system $\mathbf{A}\mathbf{x} = \mathbf{b}$ in an AC power flow problem. 

\begin{figure}[b]
    \centering
    \includegraphics[width=0.80\linewidth]{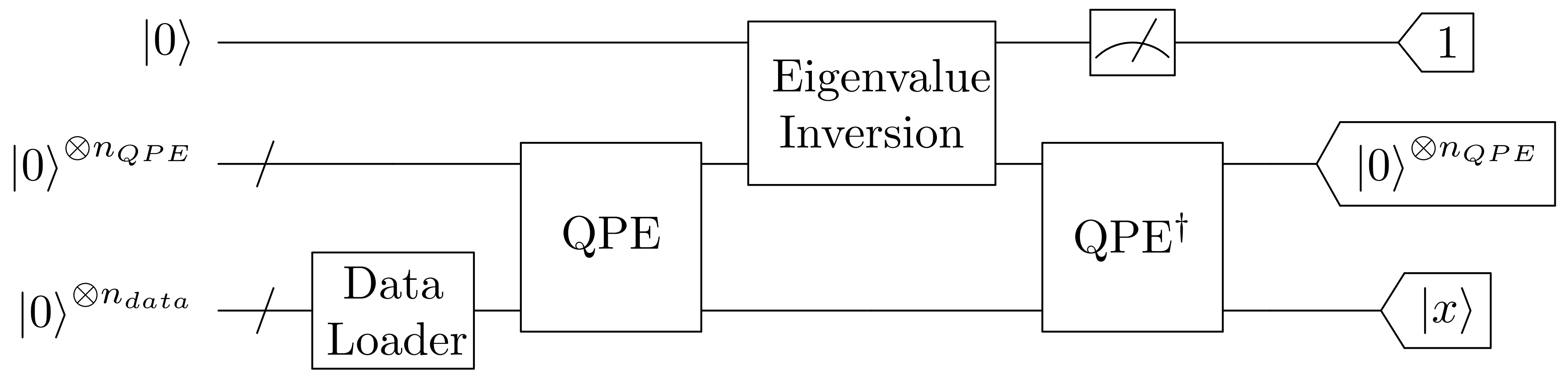}
    \caption{The conceptual circuit of HHL }
    \label{fig:hhlcirc}
\end{figure}

$\mathbf{A}$ is assumed to be Hermitian and an easy conversion can be conducted if it is not. 
HHL algorithm aims to find the normalized solution vector $|\mathbf{x}\rangle$ by
$|\mathbf{x}\rangle = \mathbf{A}^{-1}|\mathbf{b}\rangle = \sum_i \lambda_i^{-1} b_i |\mathbf{v_i}\rangle$,
where $\lambda_i$ and $\mathbf{v_i}$ are the $i^{th}$ eigenvalue and eigenvector of $\mathbf{A}$, respectively, and $|\mathbf{b}\rangle = \sum_i b_i |\mathbf{v_i}\rangle$ is the normalized vector $\mathbf{b}$ with decomposition. The decomposition of $\mathbf{b}$ in the eigenbasis of $\mathbf{A}$ is achieved with quantum phase estimation (QPE), as shown in the circuit in Fig.~\ref{fig:hhlcirc}.
The complete procedures used to solve the linear system from the power flow problems are following.

\begin{mdframed}[linecolor=black,linewidth=1pt]
\textbf{Practical HHL procedures}

\textbf{Goal:} Providing $\mathbf{A}$ and $\mathbf{b}$ in a linear system $\mathbf{A}\mathbf{x} = \mathbf{b}$, use HHL to solve for solution vector $\mathbf{x}$.

\textbf{Process:}

\begin{enumerate}
    \item First, normalize $\mathbf{b}$ to get $|\mathbf{b}\rangle$ and record $\|\mathbf{b}\|$, then check the properties of matrix $\mathbf{A}$:
    \begin{enumerate}
        \item if the condition number is too high, use a preconditioner
        \item if the dimension is not a power of 2, expand it to the nearest power of 2
        \item if the matrix is not Hermitian (e.g., due to the preconditioner), make it Hermitian:
            \begin{align*}
            \left[\begin{array}{cc}
            \mathbf{0} & \mathbf{A} \\
            \mathbf{A}^\dagger & \mathbf{0}
            \end{array}\right]
            \left[\begin{array}{c}
            \mathbf{0} \\
            |\mathbf{x}\rangle
            \end{array}\right]
            =
            \left[\begin{array}{c}
            |\mathbf{b}\rangle \\
            \mathbf{0}
            \end{array}\right].
            \end{align*}
    \end{enumerate}
    \item Generate the HHL circuit as in Figure~\ref{fig:hhlcirc}.
    \item Compute $|\mathbf{x}\rangle$ by measuring the final state $|\mathbf{\psi}\rangle$ from the circuit.
    \begin{itemize}
        \item If a simulator is used, the statevector of $|\mathbf{\psi}\rangle$ can be read directly.
        \item if a real quantum computer is used,  the statevector of $|\mathbf{\psi}\rangle$ has to be reconstructed from state tomography.
    \end{itemize}
    \item Retrieve $\mathbf{x}$ from $|\mathbf{x}\rangle$ and return it.
    \begin{enumerate}
        \item Compute 
        \begin{align*}
            \|\mathbf{x}\| = |\lambda_0| \cdot \sqrt{P(\text{measure ancilla and get 1})},
        \end{align*}
        where $|\lambda_0| \leq |\lambda_i|$ for all $i\neq0$, and $\lambda_i$ are eigenvalues of the matrix input in the circuit.
        \item Finally, $\mathbf{x} = \|\mathbf{x}\| \cdot \|\mathbf{b}\| \cdot |\mathbf{x}\rangle$.
    \end{enumerate}
\end{enumerate}
\end{mdframed}

\subsection{Resources estimation and scalability}

Following the settings in~\cite{qlsarepo} and after computing the first step in Section~\ref{sec:hhl-procedures} (expand and Hermitize the matrix $\mathbf{A}$), the estimated total number of qubits used,  $n_{\text{total}}$, is based on the following equation.
\begin{align*}
n_{\text{total}} = n_{\text{data}} + n_{\text{QPE}} + n_{\text{neg val}}
\end{align*}
where
\begin{align*}
    \centering
    \begin{cases}
        n_{\text{data}} &= \log_2(\text{number of rows or columns of matrix}) \\
        n_{\text{QPE}} &= \max(n_{\text{data}} + 1, \lceil \log_2(\kappa + 1) \rceil) \\
        n_{\text{neg val}} &= 
            \begin{cases}
                1 & \text{if the input matrix has negative eigenvalues} \\
                0 & \text{else}
            \end{cases}
    \end{cases}
\end{align*}
and $\kappa$ is the condition number of matrix $\mathbf{A}$.

Note that $n_{\text{QPE}}$ governs the accuracy of the eigenvalue computation and the solution vector $\mathbf{x}$, making it hard to reduce the qubit requirement in the QPE step, while the computational power needed in quantum simulators has an exponential dependency on $n_{\text{QPE}}$.
To mitigate this difficulty, a preconditioner is used for systems with relatively high condition numbers in our routine. The current implementation utilizes the Gauss-Seidel (GS) preconditioner due to its efficiency. 
In Section~\ref{sec:num-exp}, a case is provided to show the use of a GS preconditioner can improve the accuracy.

With NWQSim, a single NVIDIA A100 GPU can effectively handle around 32 qubits and this capacity can be expanded to 42 qubits with the utilization of multiple nodes. Considering memory space constraints, the optimal capability of solving non-singular non-Hermitian linear systems ranges from $2^{14}$-by-$2^{14}$ to $2^{20}$-by-$2^{20}$, depending on the number of 

\newpage
\noindent available GPU nodes. However, as shown in Table~\ref{tab:results}, the current bottleneck is in HHL circuit generation.
\copyrightnotice

\section{Numerical experiments\label{sec:num-exp}}

To demonstrate the capabilities of our framework, we selected the IEEE 14-bus and IEEE 30-bus systems\cite{archive}, which are readily available through the MATPOWER package. These systems serve as our initial testbed, allowing us to showcase the framework's functionality.

Experiments were conducted on Perlmutter supercomputer using a single NVIDIA A100 GPU. We will use multiple GPUs to accommodate the increased computational demands for future large test cases. In our study, HHL circuits for both the 14-bus and 30-bus cases were executed on both Qiskit Aer and the SV-Sim simulator. The detailed results are presented in Table~\ref{tab:results}. A highlight is the impact of the GS preconditioner on the 30-bus case.  It can reduce the condition number of the 30-bus system from 492.5 to 109.3. This reduction had the effect of lowering the value of $n_{\text{qpe}}$ from 10 to 8, as elaborated in the entry marked ``30-buses$^*$'' in the table.

Although GS preconditioner evidently improved the accuracy of the final solution, it came at the cost of increased circuit generation time and thus simulation time. That was because the preconditioner broke the Hermiticity of the original matrix, resulting in deeper circuits. Additionally, SV-Sim consistently outperformed Qiskit Aer, achieving simulation speeds up to 8 times faster. Notably, the running time of the C++ segment within SV-Sim was 1.7s, 36.2s, and 34.5s for the three cases, respectively. This observation indicated that the actual backend computation time for each circuit ranged from $1/150$ to $1/65$ of the time recorded in Python. This discrepancy can be attributed to the substantial time overhead involved in casting data in between C++ and Qiskit. Unfortunately, since it is difficult to obtain the actual computation time from Qiskit Aer, the authors could not make a direct comparison on simulation time without type-casting.


\section{Conclusion and future work~\label{sec:conclusion}}
This paper presents a flexible hybrid framework, offering expedited problem-solving through the incorporation of quantum algorithms while integrating with existing power system tools. The framework is based on an advanced quantum simulator that significantly minimize resource-intensive communications and GPU-based parsing through the implementation of specialized computations.
Use case with power flow studies of IEEE test systems showed up to 8 times speedup compared to Qiskit Aer with similar levels of accuracy. This framework has potential to empower power system engineers, even those without a quantum computing background, to explore and advance quantum computing and its applications in power system domain.

This paper shows an initial exploration with a focus on HHL linear algorithm in power system applications. As we look ahead, our vision includes several key improvements to further enhance the framework's capabilities. First, a C++ implementation of QPE and HHL circuit generation based on NWQSim would greatly reduce the time overhead for circuit generation and type-casting data between Qiskit and C++. With this new implementation in place, it will be feasible for running large-scale linear problems with higher number of qubits and more GPUs for scalability testing. Furthermore, authors will include more solvers into the framework, such as various variational quantum solvers, and extend it to a wide ranges of power system applications. 
Differential-algebraic equations related power system dynamics modelings have received considerable attentions as seen in~\cite{dae}. Given the underline subroutine still requires linear solvers, a worthwhile direction is to incorporated those into our framework.

\begin{table}[t]
\centering
\caption{Experiment results on 14-bus case, 30-bus case, and 30-bus case with GS preconditioner (30-bus$^*$)\label{tab:results}}
\begin{tabular}{|l|c|c|c|}\hline
                        & 14-bus  & 30-bus  & 30-bus$^*$ \\ \hline
Matrix Size             & 13 $\times$ 13     & 29 $\times$ 29     & 29 $\times$ 29                        \\ \hline
Condition Number        & 119.2     & 492.5     & 109.3                        \\ \hline
$n_{\text{total}}$ and ($n_{\text{data}}$, $n_{\text{QPE}}$)        & 13 and (4,8)        & 16 and (5,10)        & 15 and (6,8)  \\ \hline
Circuit Generation Time & 88s       & 30.5 min   & 37.5 min                      \\ \hline
Qiskit Aer Simu. Time  & 27s       & 31.3 min   & 37.5 min                      \\
 - Error $\|\mathbf{x}_{hhl} - \mathbf{x}_{true}\|_2$    & $1.97 \cdot 10^{-3}$ & $1.18 \cdot 10^{-3}$ & $6.35 \cdot 10^{-4}$   \\ \hline
SV-Sim Simu. Time  & 13s       & 3.9min   & 4.7 min                       \\
 - Error $\|\mathbf{x}_{hhl} - \mathbf{x}_{true}\|_2$             & $1.97 \cdot 10^{-3}$ & $1.18 \cdot 10^{-3}$ & $6.35 \cdot 10^{-4}$ \\ \hline
\end{tabular}
\end{table}

\end{document}